\documentclass[journal]{IEEEtran}
\usepackage{amsmath}
\usepackage{graphicx}
\usepackage{amsfonts}
\usepackage{amsmath,epsfig}
\usepackage{stfloats}
\usepackage{amssymb}
\usepackage{cite}
\usepackage{algorithm}
\usepackage{algorithmic}
\ifCLASSINFOpdf

\else

\fi

\begin{document}
%
\title{Joint Source and Relay Design for MIMO Relaying Broadcast Channels}

\author{Haibin~Wan,  Wen~Chen,~\IEEEmembership{Senior~Member,~IEEE}, and Xiaoli Wang
\thanks{Manuscript received September 26, 2012, revised October 30, 2012 and November 30, 2012; accepted
December 4, 2012. The associate editor coordinating the review of this paper and
approving it for publication was Q. Du.}
\thanks{H.~Wan, and W.~Chen are with Department of Electronic Engineering, Shanghai Jiao Tong University,
China (e-mail: \{dahai\_good; wenchen\}@sjtu.edu.cn); X.~Wang is with DoCoMo Beijing Communications Laboratories Co., Ltd (e-mail: wangxl@docomolabs-beijing.com.cn).}
\thanks{This work is supported by the National 973 Project 2012CB316106,
2009CB824904, and by NSF China 60972031 and 61161130529.}

 }


\maketitle

\begin{abstract}
In this letter, we address the optimal source and relay matrices design for the  multiple-input multiple-output~(MIMO) relaying broadcast channels~(BC)
with direct links~(DLs) based on weighted sum-rate criterion.
This problem is nonlinear nonconvex and its optimal solution remains open.
To develop an efficient way to solve this problem,
we first set up an equivalent problem, and solve it by decoupling it into two tractable subproblems. Finally, we propose a general linear iterative  design algorithm based on alternative optimization.
This iterative design algorithm is  convergent since the solution of each subproblem is optimal and unique.
The advantage of the proposed iterative design scheme is demonstrated by numerical experiments.

\end{abstract}
\IEEEpeerreviewmaketitle

\section{Introduction}
\IEEEPARstart{R}{ecently}, MIMO relaying broadcast channel~(BC) has attracted
much research interest. For a
MIMO relaying BC, there are two independent channel links
between source and receivers; i.e., \emph{source-relay-receivers } links, and
\emph{source-receivers} direct links~(DLs).  Many
works have investigated the linear strategy for MIMO relaying BC.
In~\cite{2008-Chae-fixedRelay}, an implementable multiuser precoding strategy
that combines Tomlinson-Harashima precoding at source and linear signal processing at relay is presented.
In~\cite{2009-Rui-Zhang-Qos}, a joint optimization of linear
beamforming at source and relay  to minimize  the
weighted sum-power consumption under the minimum
SINR-constraints is presented.
In~\cite{2010-WeiXu}, the singular value decomposition (SVD) and
zero  forcing~(ZF) precoder are respectively used to the source-relay channel and relay-receiver channel
to optimize the joint precoding. The authors use an iterative method
to show that the optimal precoding matrix always diagonalizes the
compound channel of the system.
In~\cite{2011-Wei-Xu}, the authors use the quadratic programming to
joint precoding optimization to maximize the system capacity.
In~\cite{2010-Gomadam}, the authors   propose a scheme based on
duality of MIMO MAC and BC to maximize
the system capacity.
All these works did not consider the DLs and each receiver is assumed to be single antenna.

In practice, the DLs provide valuable spatial diversity to the MIMO relay system and should not be ignored, especially to MIMO relaying BC. Recently, Phuyal \emph{et al.} in~\cite{2012-Phuyal} has considered the DLs
in design to deal with the power control problem, but only use the ZF scheme with single-antenna receivers.
In our previous work~\cite{2012-HBwan-WCL}, we also consider the DLs in design but only consider the RZF scheme to avoid complexity. Both schemes will cause severe noise amplification when a part of the DLs are near to zero.
In this letter, we  consider a general design strategy   to deal with the source precoding matrix~(PM) and relay beamforming matrix~(BM) by using an  alternative optimization
algorithm which is inspired by the earlier works in~\cite{2011-Shi-Luo-He,2008-Soren}. But~\cite{2011-Shi-Luo-He,2008-Soren} only focus on the classical broadcast channels without relay BM design. To incorporate relay BM with source PM will make the analysis more complicated. Finally, simulation results demonstrate that the proposed strategy outperforms the existing methods.

%
\begin{figure}[!t]
\begin{center}
\includegraphics [width=1.5in]{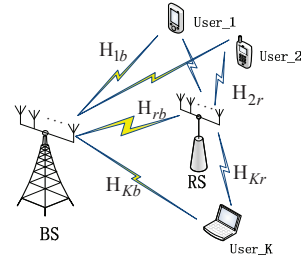}
\caption{The MIMO relaying broadcast channel  with one base station (source),
one relay station, and $K$ mobile users.} \label{Fig-Model}
\end{center}
\end{figure}

\emph{Notations:} $\textsf{E}(\cdot)$, $\mathrm{Tr}(\cdot)$, $(\cdot)^{-1}$, $(\cdot)^{T}$, $(\cdot)^{*}$, $(\cdot)^{\dag}$, and $\mathrm{det}(\cdot)$ denote expectation, trace, inverse, transpose, conjugate,  conjugate transpose, and determinant, respectively. i.i.d.  stands for independent and identically distributed. $\mathbf{I}$ is the identity matrix with appropriate dimensions. $\mathrm{diag}()$ is a diagonal matrix. $\log$ is of base $2$. $\mathcal{C}^{M\times N}$ represents the set of $M\times N$ matrices over complex field, and $\sim\mathcal{CN}(x,y)$ means satisfying  a circularly symmetric complex Gaussian distribution with  mean $x$ and covariance $y$. $\mathcal{U}=\{1,2,\cdots,K\}$.
\section{System Model}
Consider a MIMO relaying  BC where a base station~(source) is equipped with
$M$  transmit antennas to serve $K$ multiantenna users with the help of an $M$ antennas relay in a cell as depicted in
Fig~\ref{Fig-Model}.
It is also assumed that  the $k$th user is with $N_k~(\forall k\in\mathcal{U})$ receive antennas and satisfied $\sum^{K}_{k=1}N_k \leq M$ to support $N_k$ independent streams for the $k$th user
simultaneously.
We consider a two phase  transmission scheme with
a non-regenerative and half-duplex relay.

Let $\mathbf{P}\triangleq[\mathbf{P}_{1},\cdots,\mathbf{P}_{K}]$  denote the source PM where
$\mathbf{P}_{k}\in \mathcal{C}^{M\times N_{k}}$ is a PM acting on
signal vector $\mathbf{s}_{k}\sim\mathcal{CN}(\mathbf{0},\mathbf{I})$ for user $k$,
and the signal vectors for different users be independent from each other.
During the first phase, source broadcasts
the precoded data streams to relay and users  by applying a
linear PM $\mathbf{P}$.
During the second phase,  relay forwards the received signal vector
to users after a linear BM $\mathbf{F}$.
Then, the received signal vector at user $k$ can be expressed as
%
%
\begin{multline}   \label{eq:vector-yk}
\underbrace{\left[
\begin{array}{c}
           \mathbf{y}_{1k} \\
          \mathbf{y}_{2k} \\
\end{array}
       \right]}_{\mathbf{y}[k]}
=\underbrace{\left[
     \begin{array}{c}
       \mathbf{H}_{kb} \\
       \mathbf{H}_{kr}\mathbf{F}\mathbf{H}_{rb} \\
     \end{array}
   \right]}_{\mathbf{H}_{k}}\mathbf{P}_{k}\mathbf{s}_{k}+
\underbrace{\left[
                  \begin{array}{c}
                   \mathbf{n}_{1k}\\
                  \mathbf{H}_{kr}\mathbf{F}\mathbf{n}_{r}+\mathbf{n}_{2k}  \\
                  \end{array}
                \right]}_{\mathbf{G}_{k}}
\\ +
  \sum^{K}_{j=1,j\neq k}\left[
                  \begin{array}{ccc}
                   \mathbf{H}_{kb} \\
                  \mathbf{H}_{kr}\mathbf{F}\mathbf{H}_{rb}\\
                  \end{array}
                \right]
           \mathbf{P}_{j}  \mathbf{s}_{j},
\end{multline}
%
where $ \mathbf{y}_{ik}$ is the received signal during the $i$th phase at user $k$,
matrix $\mathbf{H}_{ij}$ represents the channel from the transmitter $j$ to receiver $i$, and
$\mathbf{n}_{i}~$
$\sim\mathcal{CN}(0,\mathbf{I})$ ($i=1k,~2k~\mathrm{and}~ r$) are the Gaussian
noise vectors at user $k$ during the first and second phase, and at relay, respectively.
The power constraints at source and relay
 can be expressed, respectively, as
\begin{IEEEeqnarray} {rCl}  \label{Power-c}
\sum^{K}_{k=1}\mathrm{Tr}(\mathbf{P}_{k}\mathbf{P}^{\dag}_{k})=\mathrm{Tr}(\mathbf{P}\mathbf{P}^{\dag})&\leq& {P_s}~~(~\mathrm{Source}~),\IEEEyessubnumber \label{eq:Power-source}\\
{\mathrm{Tr}(\mathbf{F}\mathbf{H}_{rb}\mathbf{P}\mathbf{P}^{\dag}\mathbf{H}_{rb}^{\dag}\mathbf{F}^{\dag} +\mathbf{FF}^{\dag})}&\leq& P_r ~~(~\mathrm{Relay}~). \IEEEyessubnumber \label{eq:Power-relay}
\end{IEEEeqnarray}
In this paper,
we treat the multi-user interference as noise and consider an MMSE  receiver at each user to deal with the received signal so that the estimated signal at user $k$ can be written as
$
\hat{\mathbf{s}}_{k}=\mathbf{A}_{k}\mathbf{y}[k].
$
According to the MMSE-filter principle~\cite{1993-EstimationTheory} and the received signal at user $k$ in (\ref{eq:vector-yk}),
the MMSE receive filter for user $k$ can be written
as:
\begin{equation}\label{eq:MMSE-receiver}
\mathbf{A}_{k}
=\mathbf{P}^{\dag}_{k}\mathbf{H}^{\dag}_{k}(\mathbf{H}_{k}\mathbf{P}\mathbf{P}^{\dag}\mathbf{H}^{\dag}_{k}+\mathbf{G}_{k}\mathbf{G}_{k}^{\dag})^{-1}
=[\mathbf{A}_{1k}~\mathbf{A}_{2k}],
\end{equation}
where $\mathbf{G}_{k}\mathbf{G}_{k}^{\dag} =
\mathrm{diag}(\mathbf{I},\mathbf{I}+
\mathbf{H}_{kr}\mathbf{FF}^{\dag}\mathbf{H}^{\dag}_{kr})$, and $\mathbf{A}_{ik}$ denotes  the receive matrix at user $k$ during the $i$th phase.

Then, the MSE-matrix for the $k$th user can be expressed as:
\begin{multline}\label{eq:MMSE-k}
\mathbf{E}_{k}=\textsf{E}\left[\|\hat{\mathbf{s}}_{k}-\mathbf{s}_{k}\|^{2}_{2} \right]
= (\mathbf{I}-\mathbf{A}_{k}\mathbf{H}_{k}\mathbf{P}_{k}) (\mathbf{I}-\mathbf{A}_{k}\mathbf{H}_{k}\mathbf{P}_{k})^{\dag}\\+
\sum_{i\neq k}\mathbf{A}_{k}\mathbf{H}_{k}\mathbf{P}_{i}\mathbf{P}^{\dag}_{i}\mathbf{H}^{\dag}_{k}     \mathbf{A}_{k}^{\dag}+\mathbf{A}_{k}\mathbf{G}_{k}\mathbf{G}^{\dag}_{k}\mathbf{A}_{k}^{\dag}.
\end{multline}
Hence, assuming Gaussian signaling for source, the achievable rate for the $k$th user during two phases is given as
\begin{IEEEeqnarray}{rCl}\label{eq:Rate-Rk}
{R}_{k}=\log\left|\mathbf{I}+\mathbf{P}^{\dag}_{k}\mathbf{H}^{\dag}_{k} \mathbf{R}^{-1}_{\mathbf{I}_{k}}\mathbf{H}_{k}\mathbf{P}_{k} \right|=\log \left|\mathbf{E}^{-1}_{k} \right|,
\end{IEEEeqnarray}
where $\mathbf{R}_{\mathbf{I}_{k}}=\mathbf{G}_{k}\mathbf{G}_{k}^{\dag}+\sum^{K}_{i=1,i\neq k}\mathbf{H}_{k}\mathbf{P}_{i}\mathbf{P}^{\dag}_{i}\mathbf{H}^{\dag}_{k}$.

The main objective of this paper is to find the PM $\mathbf{P}$ and and BM $\mathbf{F}$ to maximize the weighted sum-rate. This problem  can be expressed as the following:
\begin{equation} \label{eq:First-Problem}
\left[\mathbf{P},\mathbf{F}\right]=\arg\max_{\mathbf{P},\mathbf{F}}~ \sum^{K}_{k=1}w_{k} R_{k},~
\mathrm{s.t.:}~(\ref{Power-c}),
\end{equation}
where the weighted factor $w_{k}$   usually expresses   the  priority of the $k$th user in the system.
However, it is difficult to directly obtain the optimum closed-form solution because this optimization problem is shown to be non-linear and non-convex.

In the following sections, we first set up an equivalent problem and then propose a general iterative algorithm to design the PM $\mathbf{P}$ and  BM $\mathbf{F}$ based on  alternative optimization that updates one precoder at a time while fixing the others.

\section{Source and relay matrices Design }
To find the optimal source and relay matrices for the aforementioned problem, we first set up a new equivalent problem  by introducing
a lemma, and then we solve this new problem to replace the original problem. Furthermore, we also show two simple relay beamforming schemes with given source PM. Finally, we summarize a general iterative algorithm for source PM and relay BM.
\newtheorem{Lemma}{Lemma}
\begin{Lemma}\label{Pro:A}
Let $\mathbf{A}\triangleq \{\mathbf{A}_{k}\}^{K}_{k=1}$, and $\mathbf{W}\triangleq\{\mathbf{W}_{k}\}^{K}_{k=1}$, where $\mathbf{W}_{k}\succeq \mathbf{0}$
is a  weight matrix for the $k$th receiver. Then, the optimal solution for the following problem is also the solution for the original  problem formulated in (\ref{eq:First-Problem}):
\begin{equation} \label{eq:Min-Problem}
\min_{\mathbf{P},\mathbf{F},\mathbf{W},\mathbf{A}}
\sum^{K}_{k=1}w_{k}\left(\mathrm{Tr}(\mathbf{W}_{k}\mathbf{E}_{k})-\log\det(\mathbf{W}_{k}) \right),\mathrm{s.t.:}(\ref{Power-c}),
\end{equation}
\end{Lemma}
\begin{IEEEproof}
The  proof is similar to \emph{Theorem~1} in~\cite{2011-Shi-Luo-He} and   omitted to save space.
\end{IEEEproof}

In fact, this Weighted MMSE (WMMSE) problem is also non-linear and non-convex. But, if fixing three of the four variables, the problem is  convex with respect to (w.r.t.) the remaining  variable,  and the solution has a closed-form.
In particular, with given $\mathbf{P}$, $\mathbf{F}$  and $\mathbf{A}$, the optimal weight matrix $\mathbf{W}_{k}$ can be expressed in a closed-form as
\begin{IEEEeqnarray} {rCl}\label{eq:Wk}
\mathbf{W}_{k}=\mathbf{E}^{-1}_{k},
\end{IEEEeqnarray}
and the solution for MMSE-receiver $\mathbf{A}_{k}$ is given in~(\ref{eq:MMSE-receiver}).
\subsection{Source Precoding Matrix Design}
If the relay BM $\mathbf{F}$ is given,  the MIMO relaying BC becomes  a general MIMO BC~\cite{2008-Soren,2011-Shi-Luo-He}.
Thus, for given $\mathbf{F}$,  $\mathbf{W}$ and $\mathbf{A}$, the problem in (\ref{eq:Min-Problem}) w.r.t. $\mathbf{P}$ can be reformulated as
\begin{equation} \label{eq:Min-Problem-P}
\min_{\mathbf{P}}
\sum^{K}_{k=1}w_{k}\left(\mathrm{Tr}(\mathbf{W}_{k}\mathbf{E}_{k})\right),
~~~~\mathrm{s.t.:}~~(\ref{eq:Power-source})\footnote{Here, the relay power constraint is ignored which does not affect the final result since it will be dealt with the iterative algorithm.},
\end{equation}
which is a convex quadratic optimization problem and can be solved by KKT conditions.
Thus,
we can readily obtain the Lagrangian function of (\ref{eq:Min-Problem-P}) as
\begin{IEEEeqnarray} {rCl}\label{eq:Lagran-Problem}
&&\mathcal{L}(\{\mathbf{P}_{k}\}^{K}_{k=1})=
\sum^{K}_{k=1}\Bigg(w_{k} \left(\mathrm{Tr}\big(\mathbf{W}_{k}(\mathbf{I}-\mathbf{\widetilde{H}}_{k}\mathbf{P}_{k}) (\mathbf{I}-\mathbf{\widetilde{H}}_{k}\mathbf{P}_{k})^{\dag}\big) \right)
\nonumber\\&&
+\sum_{j\neq k}w_{j}\left( \mathrm{Tr}\left(\mathbf{W}_{j}\mathbf{\widetilde{H}}_{j}\mathbf{P}_{k}\mathbf{P}^{\dag}_{k}\mathbf{\widetilde{H}}^{\dag}_{j}     \right)\right) \Bigg)
+
\lambda \left(\mathrm{Tr}\left(\mathbf{P}\mathbf{P}^{\dag}\right)- P_s\right).\nonumber
\end{IEEEeqnarray}
where $\mathbf{\widetilde{H}}_{i}=\mathbf{A}_{i} \mathbf{H}_{i}$.
Then, the  first-order necessary  condition of $\mathcal{L}$ w.r.t. each $\mathbf{P}_{k}$ yields 
\begin{equation} \label{eq:P-solution}
\mathbf{P}_{k}(\lambda)=w_{k} \left(\sum^{K}_{j=1} w_{j} \mathbf{\widetilde{H}}^{\dag}_{j}\mathbf{W}_{j} \mathbf{\widetilde{H}}_{j} +\lambda\mathbf{I}  \right)^{-1}
\mathbf{\widetilde{H}}^{\dag}_{k}  \mathbf{W}_{k} ,
\end{equation}
where $\lambda\geq 0$ is the Lagrangian multiplier which should satisfy the KKT complementarity conditions for power budget constraint, i.e., $\lambda \left(\mathrm{Tr}\left(\mathbf{P}\mathbf{P}^{\dag} \right)- P_s\right)=0$ and $\mathrm{Tr}\left(\mathbf{P}\mathbf{P}^{\dag} \right)\leq P_s$.
The $\lambda$ can be found by a 1-D search method such as the bisection method since $\mathrm{Tr}(\mathbf{P}(\lambda)\mathbf{P}(\lambda)^{\dag})$ is monotonically decreasing function of $\lambda$.

\subsection{ Relay Beamforming Matrix Design}
To find an optimal relay BM $\mathbf{F}$ for the original problem, we also consider dealing with the new problem in (\ref{eq:Min-Problem}). Thus,  with given $\mathbf{P}$, $\mathbf{W}$ and $\mathbf{A}$, the problem in (\ref{eq:Min-Problem}) w.r.t.
relay BM $\mathbf{F}$ can be recast as following
\begin{equation} \label{eq:SumMSE-F}
\min_{\mathbf{F}}~~
\sum^{K}_{k=1} w_{k}\left(\mathrm{Tr}(\mathbf{W}_{k}\mathbf{E}_{k}) \right),~~
\mathrm{s.t.:}~~(\ref{eq:Power-relay}).
\end{equation}
This WMMSE problem w.r.t. $\mathbf{F}$ can be proven to be convex by the following lemma.
\begin{Lemma}\label{Pro:B}
With given $\mathbf{P}$,
$\mathbf{A}$ and  $\mathbf{W}$, the problem of relay BM design  to minimize the total weighted-MSE in the  considered  relaying BC formulated in (\ref{eq:SumMSE-F})
is convex.
\end{Lemma}
\begin{IEEEproof}
The  proof is similar to Appendix~A in~\cite{2012-RuiWang-Tao} and   omitted to save space.
\end{IEEEproof}
\newtheorem{Remark}{Remark}
\begin{Remark}
 Here, we have a similar argument as in~\cite{2012-RuiWang-Tao} which is to deal with the two-way relaying channel with only two transceivers and a relay node based on MSE criterion.
\end{Remark}

Thus, the Lagrangian function of (\ref{eq:SumMSE-F}) for $\mathbf{F}$ is given as
\begin{IEEEeqnarray} {rCl}\label{eq:Lagran-Problem-F}
&&\mathcal{L}(\mathbf{F})=
\sum^{K}_{k=1}\Bigg(w_{k} \left(\mathrm{Tr}\big(\mathbf{W}_{k}(\mathbf{I}-\mathbf{A}_{k}\mathbf{H}_{k}\mathbf{P}_{k}) (\mathbf{I}-\mathbf{A}_{k}\mathbf{H}_{k}\mathbf{P}_{k})^{\dag}\big) \right)
\nonumber\\&& \left.
+w_{k}\mathrm{Tr}\left(\mathbf{W}_{k}(\sum_{i\neq k}\mathbf{A}_{k}\mathbf{H}_{k}\mathbf{P}_{i}\mathbf{P}^{\dag}_{i}\mathbf{H}^{\dag}_{k}     \mathbf{A}_{k}^{\dag}+\mathbf{A}_{k}\mathbf{G}_{k}\mathbf{G}^{\dag}_{k}\mathbf{A}_{k}^{\dag})\right)
 \right)
\nonumber\\&&
+\mu \left(\mathrm{Tr}(\mathbf{F}\mathbf{H}_{rb}\mathbf{P}\mathbf{P}^{\dag}\mathbf{H}_{rb}^{\dag}\mathbf{F}^{\dag} +\mathbf{FF}^{\dag})-P_r\right).
\end{IEEEeqnarray}
Before dealing with the KKT conditions, we first substitute $\mathbf{A}_{k}\triangleq[\mathbf{A}_{1k}~\mathbf{A}_{2k}]$ and $\mathbf{H}_{k}\triangleq\left[
\begin{array}{c}
\mathbf{H}_{kb} \\
\mathbf{H}_{kr}\mathbf{F}\mathbf{H}_{rb} \\
\end{array}
\right] $ into (\ref{eq:Lagran-Problem-F}) to get a function w.r.t. $\mathbf{F}$.
Then, the KKT conditions can be expressed as following
%
\begin{equation} \label{eq:KKT-F*}
\frac{\partial f}{\partial\mathbf{F}^{*}}=
\sum^{K}\limits_{k=1} w_{k}\mathbf{\Delta}_{k}
+
\left(\sum^{K}\limits_{k=1} w_{k}
\mathbf{\Theta}_{k} +\mu\mathbf{I}\right)
\mathbf{F}\left(\mathbf{\Pi}+\mathbf{I}\right)=0,
\end{equation}
\begin{IEEEeqnarray} {rCl}
\mu \left(\mathrm{Tr}\left(\mathbf{F}(\mathbf{\Pi}+\mathbf{I})\mathbf{F}^{\dag}\right)-P_r\right) &=&0,
\\
\mathrm{Tr}\left(\mathbf{F}(\mathbf{\Pi} +\mathbf{I})\mathbf{F}^{\dag}\right)&\leq& P_r,
\end{IEEEeqnarray}
%
%
where  $\mathbf{\Pi}\triangleq \mathbf{H}_{rb}\mathbf{P}\mathbf{P}^{\dag}\mathbf{H}_{rb}^{\dag}$, $\mathbf{\Theta}_{k}\triangleq\mathbf{H}^{\dag}_{kr}\mathbf{A}^{\dag}_{2k}\mathbf{W}_{k}\mathbf{A}_{2k}\mathbf{H}_{kr} $, and
$\mathbf{\Delta}_{k}\triangleq \mathbf{H}^{\dag}_{kr}\mathbf{A}^{\dag}_{2k}\mathbf{W}_{k}\mathbf{A}_{1k}\mathbf{H}_{kb}\mathbf{P}\mathbf{P}^{\dag}\mathbf{H}_{rb}^{\dag}
-\mathbf{H}^{\dag}_{kr}\mathbf{A}^{\dag}_{2k}\mathbf{W}_{k}\mathbf{P}^{\dag}_{k}\mathbf{H}^{\dag}_{rb}
$.
Based on (\ref{eq:KKT-F*}), we can obtain
\begin{equation} \label{eq:F-solution}
\mathbf{F}=
\left(\sum^{K}\limits_{k=1} w_{k}
\mathbf{\Theta}_{k} +\mu\mathbf{I}\right)^{-1}
\left(\sum^{K}\limits_{k=1}-w_{k}\mathbf{\Delta}_{k}\right)(\mathbf{\Pi} +\mathbf{I})^{-1},
\end{equation}
where $\mu$ is the Lagrangian multiplier which  can also be solved by a 1-D search method since $\mathrm{Tr}(\mathbf{F}(\mu)(\mathbf{\Pi}+\mathbf{I})\mathbf{F}(\mu)^{\dag})$ is monotonically decreasing function of $\mu$.
\subsection{ Relay Beamforming Matrix Design by Other Schemes }
If the source PM $\mathbf{P}$ is given, the relay BM can be also constructed by following schemes.

\subsubsection{MRC and MRT Beamforming }
According to the principles of maximum ratio combining (MRC) and maximum ratio transmission (MRT), we can formulate the MRC-MRT relay beamforming scheme for given source PM $\mathbf{P}$ as
\begin{IEEEeqnarray}{rCl} \label{eq:MRC-MRT}
\mathbf{F}=\rho_{1}\mathbf{H}^{\dag}_{ur}\mathbf{P}^{\dag}\mathbf{H}^{\dag}_{rb},~(\mathbf{H}_{ur}\triangleq [\mathbf{H}^{T}_{1r},\cdots,\mathbf{H}^{T}_{Kr}]^{T})
\end{IEEEeqnarray}
where the factor  $\rho_{1}$ can be derived from (\ref{eq:Power-relay}).
\subsubsection{MRC and RZF Beamforming }
%
According to the principles of MRC and regularized zero-forcing transmission (RZF)~\cite{2009-Sung-GCI}, we can formulate the MRC-RZF relay beamforming scheme for given source PM $\mathbf{P}$ as
\begin{IEEEeqnarray}{rCl} \label{eq:MRC-MRT}
\mathbf{F}=\rho_{2}\mathbf{H}^{\dag}_{ur}(\mathbf{H}_{ur}\mathbf{H}^{\dag}_{ur}+\frac{M}{P_r}\mathbf{I})^{-1}\mathbf{P}^{\dag}\mathbf{H}^{\dag}_{rb},
\end{IEEEeqnarray}
where the factor  $\rho_{2}$ can also be derived from (\ref{eq:Power-relay}).
\subsection{A General Iterative Design Algorithm} \label{sec: Algorithm}
In summary, a general design algorithm for source PM $\mathbf{P}$ and relay BM $\mathbf{F}$ can be summarized as following:
\begin{algorithm} \label{Algorithm-1}
\caption{\textbf{:}~A General Iterative Design Algorithm}
 \begin{algorithmic}[1]
 \STATE
\textbf{Initialize:} $\mathbf{P}=\sqrt{\frac{P_s}{M}}\mathbf{I}$, $\mathbf{F}=\rho\mathbf{I}$,  $\mathbf{A}_{k}~(\forall k\in \mathcal{U})$ uses (\ref{eq:MMSE-receiver}) with $\mathbf{P}=\sqrt{\frac{P_s}{M}}\mathbf{I}$ and $\mathbf{F}=\rho\mathbf{I}$, where $\rho$ satisfies the power constraints.
 \STATE {\textbf{Repeat:}}
 \STATE Update $\mathbf{P}_{k}$ using (\ref{eq:P-solution}) for fixed $\mathbf{W}_{k}$, $\mathbf{A}_{k}$ and $\mathbf{F}$, $\forall k\in \mathcal{U}$;
 \STATE Update $\mathbf{F}$ using (\ref{eq:F-solution}) for fixed $\mathbf{W}_{k}$, $\mathbf{A}_{k}$ and $\mathbf{P}$, $\forall k\in \mathcal{U}$;
\STATE Update $\mathbf{A}_{k}$ using (\ref{eq:MMSE-receiver})  for fixed  $\mathbf{P}$ and $\mathbf{F}$, $\forall k\in \mathcal{U}$;
\STATE Update $\mathbf{W}_{k}$ using (\ref{eq:MMSE-k}) and (\ref{eq:Wk}) for fixed  $\mathbf{P}$ and $\mathbf{F}$, $\forall k\in \mathcal{U}$;
 \STATE \textbf{Until:} The termination criterion is satisfied.
\end{algorithmic}
\end{algorithm}

This  algorithm is always convergent  to a stationary point.
The convergence analysis of this algorithm is similar to  Theorem~2 in~\cite{2012-RuiWang-Tao} or referred to the block coordinate descent algorithm in~\cite{1999-NLP-Book}. Simulation part will show the convergence.

\section{Numerical Result}
This section presents numerical results to evaluate the proposed precoding designs over $2000$
random channel realizations.
 For fair comparison,
the other schemes  for  comparison are also considering the DLs contributions,
which are: 1) BZF-BZF\&BZF~in~\cite{2012-Phuyal}, 2) BRZF-BZF\&BRZF~in~\cite{2012-HBwan-WCL}, and 3) SVD-RZF~in~\cite{2011-WZJ}. All these schemes are adjusted to suitable for the multi-antenna users case as in~\cite{2009-Sung-GCI}  for fair comparison.
The channel gains  are set to be the combination of
large scale fading and  small scale
fading,
 i.e., all channel matrices have
i.i.d.  $\mathcal{CN}(0,\frac{1}{\ell^{\tau}})$ entries, where
$\ell$ is the distance between two nodes, and $\tau=3$ is the path loss
exponent.
In these simulations, we
consider  that  BS and relay are deployed in a line with users,
where all the users are deployed at the same point.

\begin{figure}[!t]
\begin{center}
\includegraphics [width=2.4in]{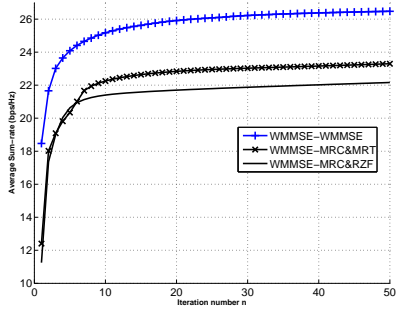}
\caption{Convergence properties for 1 (randomly selected) channel realization with
$P_s=P_r (\mathrm{SNR}=28 \mathrm{dB})$, where  $M=6, K=3$,  BS is at 0 point, relay is at 0.5 point, all users are at 1.0 point, and $w_{k}=1$, $\forall k\in \mathcal{U}$.} \label{Fig-Iterative}
\end{center}
\end{figure}

\begin{figure}[!t]
\begin{center}
\includegraphics [width=2.4in]{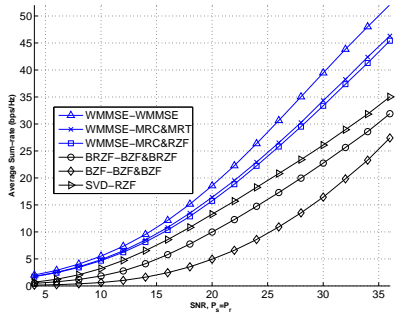}
\caption{Average sum-rate versus the transmit power with
$P_s=P_r$ (SNR dB), where  $M=8,~K=4$,  BS is at 0 point, relay is at 0.5
point, all users are at 1.0 point, and $w_{k}=1$, $\forall k\in \mathcal{U}$. } \label{Fig-SNR}
\end{center}
\end{figure}

\begin{figure}[!t]
\begin{center}
\includegraphics [width=2.4in]{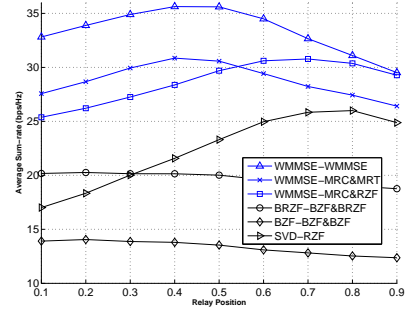}
\caption{Average sum-rate versus relay's position~(between 0 and 1.0 ),
where BS is at $0$ point and all users are at $1.0$ point,
$M=8, K=4$, $P_s=P_r~ (\mathrm{SNR}=28\mathrm{dB}$), and $w_{k}=1$, $\forall k\in \mathcal{U}$.
 } \label{Fig-Position}
\end{center}
\end{figure}

We first show the  convergence properties of the proposed precoding strategy in Fig.~\ref{Fig-Iterative}.
Fig.~\ref{Fig-SNR} shows  the average sum-rate of the network versus the transmitting power, when all nodes positions are fixed.
Fig.~\ref{Fig-Position} shows the average sum-rate of the network versus the
relay's position, when the powers at BS and relay are fixed.
From Fig.~\ref{Fig-SNR}  and Fig.~\ref{Fig-Position},  we
can see that the average sum-rate of the proposed strategy is
higher than those of the other linear schemes at all SNR regime and all relay's position.
This is because that the DLs contributions of SVD-RZF scheme are approximately  equal to zero, and the BZF-BZF\&BZF  and BRZF-BZF\&BRZF schemes  will  amplify noise signal, especially at
the case that the DLs gains are close to zero. However, the proposed WMMSE-WMMSE scheme can better deal with  the multi-user interferences and noise, and the relation between DLs gains and source-relay-users channel gains.

\section{Conclusion }
In this letter,  we consider the joint source and relay matrices design for the MIMO relaying BC with considering the source-receivers DLs based on sum-rate criterion.
To deal with  the source-relay-receivers channels and DLs gains, and the multi-user interference, we propose a general  joint design  strategy based on alternative optimization algorithm in which we solve an equivalent problem to replace the original problem by a WMMSE method.
We also present other two linear relay beamforming schemes with less complexity.
Numerical results show that  the proposed strategy  outperforms
the other linear schemes with or without considering DLs in design.





\bibliography{mybib}
\bibliographystyle{ieeetr}
%


%

\end{document}